\documentclass[twocolumn, 10pt, pre, aps, showpacs, reprint, amsmath, amssymb, superscriptaddress, nofootinbib]{revtex4-2}
\usepackage{amsmath}
\usepackage{graphicx}
\begin{document}

\title{Polarization and tranverse mode nonlinear dynamics in a multimode VCSEL}
\author{Yohann G. Sanvert}
\email{yohann.sanvert@centralesupelec.fr}
\author{Jules Mercadier}
\author{Stefan Bittner}
\affiliation{Universit\'e de Lorraine, CentraleSup\'elec, LMOPS EA-4423, 57070 Metz, France}
\affiliation{Chaire Photonique, LMOPS EA-4423, CentraleSup\'elec, 57070 Metz, France}
\author{Angel Valle}
\affiliation{Instituto de Física de Cantabria (CSIC-UC), Facultad de Ciencias Avda. Los Castros s/n, E-39005 Santander, Spain}
\author{Marc Sciamanna}
\email{marc.sciamanna@centralesupelec.fr}
\affiliation{Universit\'e de Lorraine, CentraleSup\'elec, LMOPS EA-4423, 57070 Metz, France}
\affiliation{Chaire Photonique, LMOPS EA-4423, CentraleSup\'elec, 57070 Metz, France}

\begin{abstract}
We theoretically analyze the nonlinear dynamics and routes to chaos in a multimode vertical cavity surface-emitting laser (MM-VCSEL) in free-running operation. Including higher order transverse modes (TMs) results in additional bifurcations at higher currents not found for single-mode VCSELs (SM-VCSELs). The resulting dynamics involve competition between modes with different transverse profiles and polarization and show good qualitative agreement with recent experiments.
\end{abstract}

\maketitle

The intrinsic polarization competition in VCSELs enables the generation of highly complex chaos in free-running operation \cite{virte2013deterministic,virte2014physical}, i.e., without optical injection or feedback \cite{virte2014physical, law1997effects, panajotov2012optical}. Polarization chaos has been successfully used for random number generation or chaos cryptography \cite{hong2004synchronization, wang2024chaos}. More recently, a close inspection into the dynamics of multimode VCSELs revealed many parameter regions with chaotic dynamics \cite{Bittner2022, mercadier2025chaos}. Chaos is typically characterized by a strong competition between transverse modes with different polarizations and polarization switchings (PSs). While there exist many theoretical studies of the nonlinear dynamics of SM-VCSELs \cite{san1995light,erneux1999two,virte2013bifurcation,virte2014bistability}, only few theoretical studies have handled the case of MM-VCSELs and have mostly concentrated on static, not dynamic, properties \cite{Law1997a, MartinRegalado1997, valle1998polarization, Rimoldi2025}. % paragraph

Here, we extend earlier analyses of MM-VCSEL dynamics. While we restrict the model to only two TMs, it captures the main features of recent experiments \cite{Bittner2022, mercadier2025chaos}. This restriction is adopted to study the simplest possible case of multimode dynamics and to see if even a single additional TM significantly impacts the dynamics. Several physical effects must be considered in a MM-VCSEL model. The output field has two linear polarizations, $x$ and $y$, and the related PSs are modeled via the spin-flip model (SFM) \cite{san1995light}. Due to birefringence induced during fabrication, the frequencies of $x$- and $y$-polarized modes are different. The inhomogeneous spatial intensity distribution in the VCSEL leads to spatial hole burning (SHB), i.e., the creation of "holes" in the carrier density where high optical intensity causes more carriers to recombine. 
SHB together with spin-carrier coupling is proposed to explain the polarization properties \cite{valle1995spatial}. % paragraph

% figure 1
\begin{figure*}[tb] 
    \centering
    \centering
    \includegraphics[width=\linewidth]{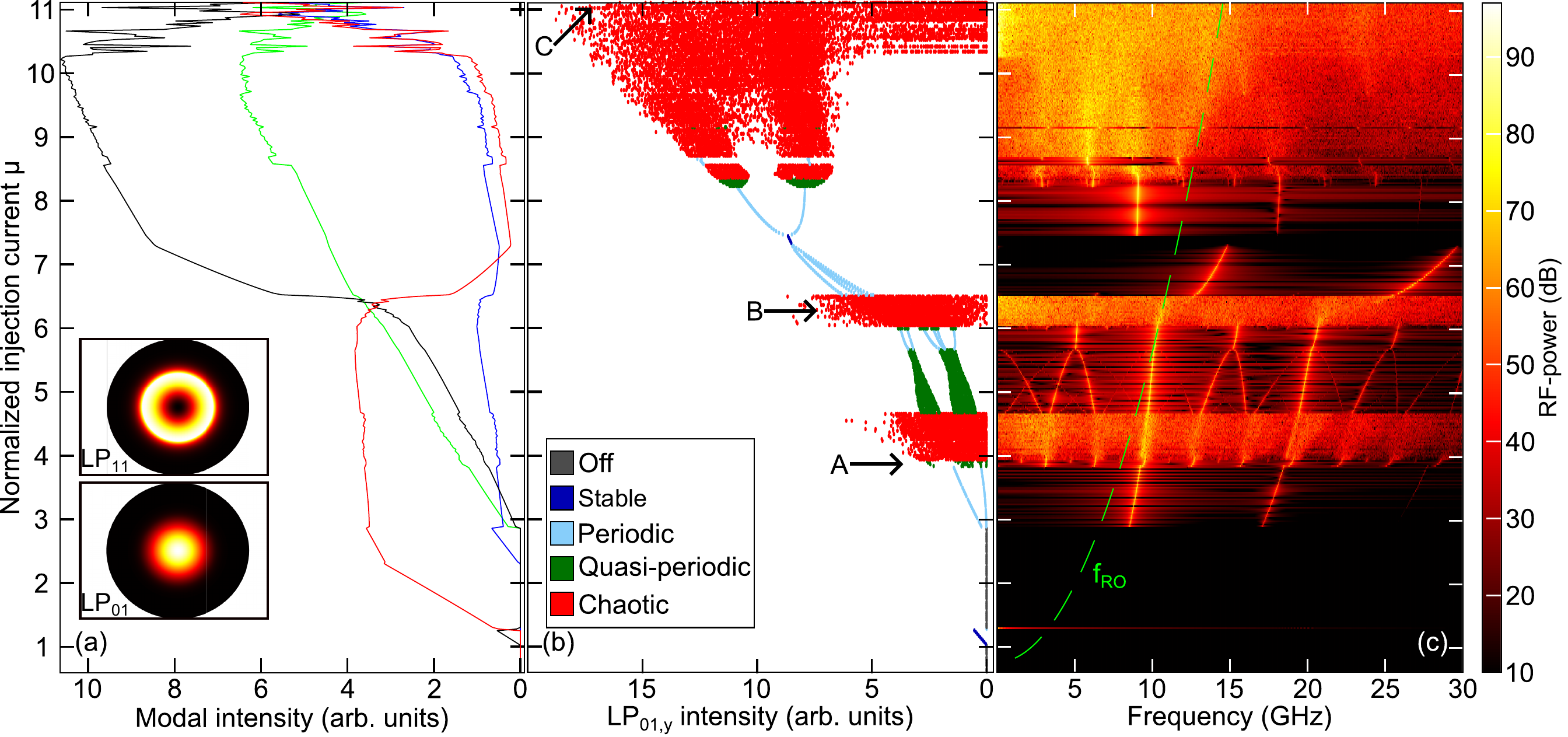}
    \caption{(a) LI-curves for the LP$_{01,x}$ (red), LP$_{01,y}$ (black), LP$_{11,x}$ (green) and LP$_{11,y}$ (blue) modes for $f_{p,0}$ = 9.7 GHz and $\gamma_{a}$ = 1 ns$^{-1}$. Inset: 2D intensity profiles of the modes LP$_{01}$ and LP$_{11}$ (superposition of both orientations). (b) Bifurcation diagram and (c) RF-spectrum of the LP$_{01,y}$ mode. The green dashed line is the ROF. The three points A, B and C in (b) correspond to Fig.~\ref{fig:trace_spectre_quasi_per_chaos}(A), Fig.~\ref{fig:trace_spectre_quasi_per_chaos}(B) and Fig.~\ref{fig:polarization_transverse_modes_switching}, respectively.}
    \label{fig:diag_bifur_spectre_RF}
\end{figure*}

In this article, a circular active region with radius $a = 3~\mu$m is used. While in experiments \cite{Bittner2022, mercadier2025chaos} up to twelve TMs lase, only the fundamental (LP$_{01}$) and the first TM (LP$_{11}$) are considered here, where the index $j \in \{0, 1\}$ denotes the LP$_{j1}$ mode. Then, the total optical field may be written in the linear basis with unit vectors \mbox{$\vec{x}$ and $\vec{y}$ as \cite{valle1998polarization}}
\begin{align}
    \nonumber
    &\vec{E}(t,r) = \{[E_{0,x}(t)\psi_{0,x}(r) + E_{1,x}(t)\psi_{1,x}(r)]\vec{x}\\
    &+ [E_{0,y}(t)\psi_{0,y}(r) + E_{1,y}(t)\psi_{1,y}(r)]\vec{y}\}e^{i\kappa\alpha t} + \mathrm{c.c.}
    \label{E_tot}
\end{align}
where $E_{ji}$ and $\psi_{ji}$ with $j \in\{0,1\}$ are respectively the amplitude and the profile of the TM LP$_{j1}$ with polarization $i\in\{x,y\}$. The SFM model extended for MM-VCSELs reads \cite{valle1998polarization, valle1995spatial}
\begin{align}		
    \nonumber
    \dot{E}_{0x,0y}(t) = & \kappa(1+i\alpha)[(g_{0x,0y}-1)E_{0x,0y} \pm i g_{0xy,0yx}E_{0y,0x}] \\ 
		 & \mp (\gamma_{a}+i\gamma_{p,0})E_{0x,0y} + F_{0x,0y}		
    \label{E_0_dot} \\ \nonumber \\
		\nonumber
    \dot{E}_{1x,1y}(t) = & \kappa(1+i\alpha)[(g_{1x,1y}-1)E_{1x,1y} \pm ig_{1xy,1yx}E_{1y,1x}] \\ 
		& \mp (\gamma_{a}+i\gamma_{p,1})E_{1x,1y} + i \gamma_{p}^{(tr)}E_{1x,1y} + F_{1x,1y}
		\label{E_1_dot} \\ \nonumber \\
    \nonumber
    \dot{N}(t, r) = & J(t,r) +D\nabla_{\perp}^2 N - \gamma_{e}[N(1 + \sum_{i=x,y}{\sum_{j=0,1}{|E_{ji}|^{2}\psi_{ji}^{2}}}) \\
    &-in\sum_{j=0,1}{(E_{jx}E_{jy}^{*}-E_{jx}^{*}E_{jy})\psi_{jx}\psi_{jy}}]
    \label{N_dot} \\ \nonumber \\
    \nonumber
    \dot{n}(t, r) = & -\gamma_{s}n +D\nabla_{\perp}^2 n - \gamma_{e}[n\sum_{i=x,y}{\sum_{j=0,1}{|E_{ji}|^{2}\psi_{ji}^{2}}} \\
    &-iN\sum_{j=0,1}{(E_{jx}E_{jy}^{*}-E_{jx}^{*}E_{jy})\psi_{jx}\psi_{jy}}]
    \label{n_dot}
\end{align}
where a dot denotes the time derivative, and $N$ ($n$) is the total (difference) of carrier inversions with opposite spins. $N$, $n$ and the current density $J(t, r)$ are multiplied with the differential gain and divided by the cavity loss rate as normalization~\cite{san1995light}. The intra- and inter-modal gains characterizing the overlap of the carrier density and the transverse mode profiles are  
\begin{equation} \label{g_ijk} \begin{array}{rcl} g_{ji} & = & \frac{\int_{0}^{\infty}{N(r)\psi_{ji}^{2}(r)rdr}}{\int_{0}^{\infty}{\psi_{ji}^{2}(r)rdr}} \, , \\ \\
g_{jik} & = & \frac{\int_{0}^{\infty}{n(r)\psi_{ji}(r)\psi_{jk}(r)rdr}}{\int_{0}^{\infty}{\psi_{ji}^{2}(r)rdr}} \end{array}
\end{equation}
with $j \in\{0,1\}$ and $i, k\in\{x, y\}$ where $i \neq k$. The explicit spatial dependence of $N$, $n$ and the gains allows to include SHB and carrier diffusion effects. % paragraph

The constants used in Eqs.~(\ref{E_tot}-\ref{n_dot}) are the field decay rate $\kappa$ = 300 ns$^{-1}$, the linewidth enhancement factor $\alpha$ = 3, the amplitude anisotropy $\gamma_{a}$, the birefringence-induced frequency splittings $f_{p,j} = \gamma_{p,j}/\pi$ between the LP$_{j1,x}$ and LP$_{j1,y}$ modes, which have almost the same values, the frequency splitting $f_{p}^{(tr)} = \gamma_{p}^{(tr)}/(2\pi) \approx$ 162~GHz between the LP$_{01,x}$ and LP$_{11,x}$ modes, the injection current $J(t, r)$, the diffusion constant $D$ = 10 cm$^2$/s, the carrier decay rate $\gamma_{e}$ = 1 ns$^{-1}$ and the spin relaxation rate $\gamma_{s}$ = 50 ns$^{-1}$. Spontaneous emission noise is modeled through	the terms $F_{jx} = F_j$ and $F_{jy} = -i F_j$ with $F_j = \sqrt{\beta/2}(\sqrt{\bar{N}-\bar{n}} \, \xi_{j,-}+\sqrt{\bar{N}+\bar{n}} \, \xi_{j,+})$ where $\xi_{j,\pm}$ are independent zero-mean Gaussian random variables, $\bar{N}$ ($\bar{n}$) is the spatial averages over the active region of $N$ ($n$), and $\beta = 0.1$ is the spontaneous emission factor. Simulations without noise show similar results, the dynamical scenarios reported here are therefore of deterministic origin. % paragraph

To simulate this set of equations, a finite-difference time-domain (FDTD) method was employed, using an explicit Euler integration scheme with a temporal (spatial) step of $0.01$~ps ($0.12~\mu$m). For each current value, the system was simulated for $50$~ns. Furthermore, when increasing the current, the final state for the preceding current was used as initial conditions. We calculate the birefringence splittings $f_{p,j}$ by solving the Helmholtz equation for each TM, imposing the continuity of the tangential field component and its normal derivative at the interface of active region and cladding \cite{valle1995spatial}. % paragraph

Light-intensity (LI) curves of the modes are plotted in Fig.~\ref{fig:diag_bifur_spectre_RF}(a) for $f_{p,0} \approx 9.7$~GHz and $\gamma_{a}$ = 1 ns$^{-1}$. They are calculated from the average intensity over the last $20$~ns for each value of the normalized injection current \mbox{$\mu = J(r = 0,t) / J_{th}$} where $J_{th} \simeq 1.35$ is the threshold current. The inset displays the 2D profiles of the LP$_{01}$ and LP$_{11}$ modes. For $\mu \leq 2.6$, the system exhibits a behavior similar to a SM-VCSEL \cite{martin1996polarization}. Indeed, the LP$_{11}$ mode emerges only above $\mu = 2.6$, hence the multimode model reproduces a single-mode model for low currents. A PS between the LP$_{01,x}$ and LP$_{01,y}$ modes is observed at $\mu \approx 1.2$. Just after the onset of the LP$_{11}$ modes, the behavior becomes more complex as the mode hierarchy is rearranged, with another PS at $\mu \approx 6.5$. Lastly, for very high currents, the LI-curves show significant fluctuations that suggest complex dynamics including redistribution of power between both transverse and polarization modes. % paragraph

\begin{figure}[t] 
    \centering
    \includegraphics[width=\linewidth]{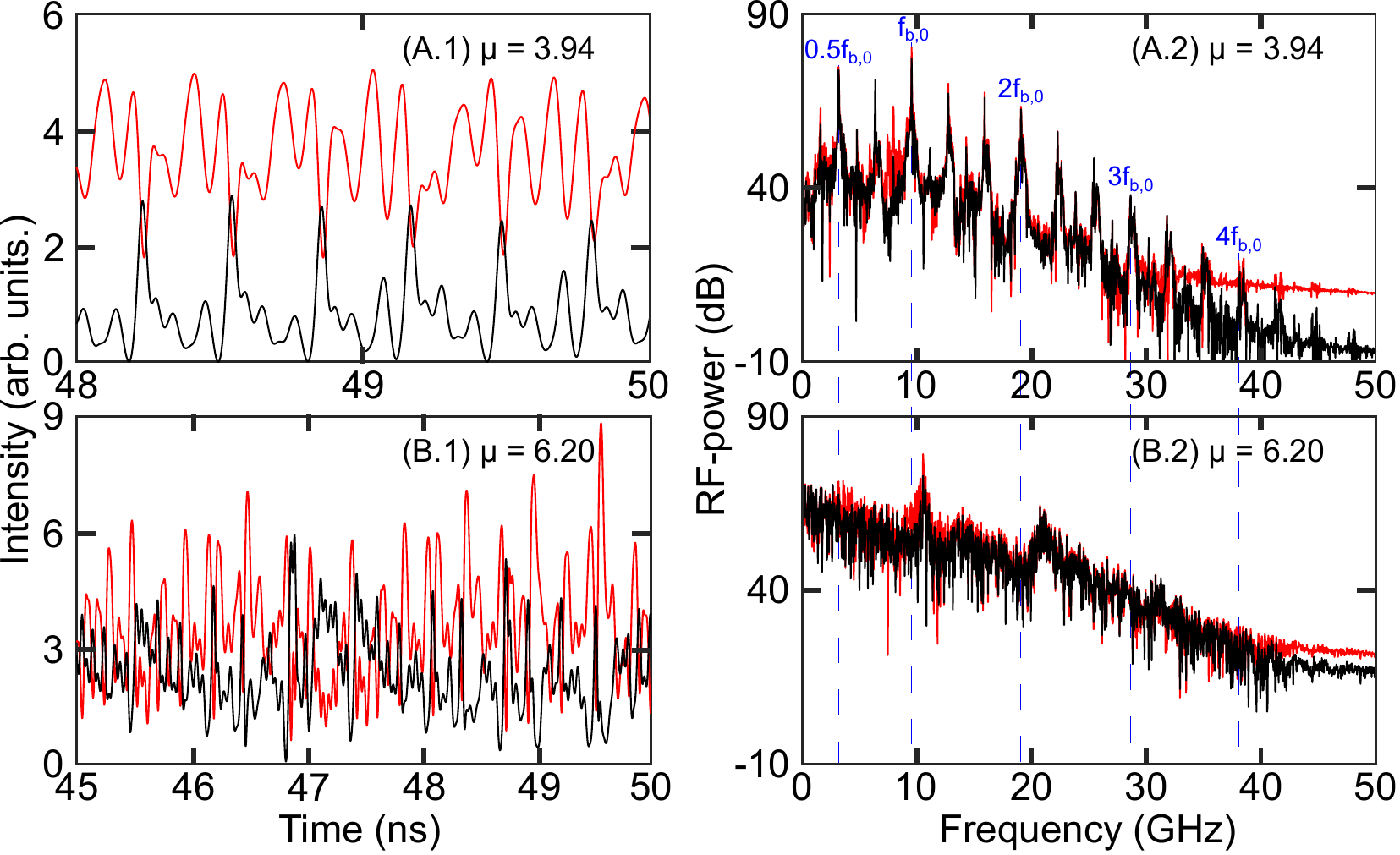}
    \caption{(A.1) Temporal trace and (A.2) RF-spectrum for $\mu = 3.94$ and (B) for $\mu = 6.20$ with $f_{p,0} = 9.7$~GHz and $\gamma_a = 1~\mathrm{ns}^{-1}$. Mode LP$_{01,x}$ (LP$_{01,y}$) is shown in red (black). The birefringence splitting $f_{p,0}$ and its harmonics are indicated in blue.}
    \label{fig:trace_spectre_quasi_per_chaos}
\end{figure}

\begin{figure*}[t!] % figure 4
    \centering
    \includegraphics[width=\linewidth]{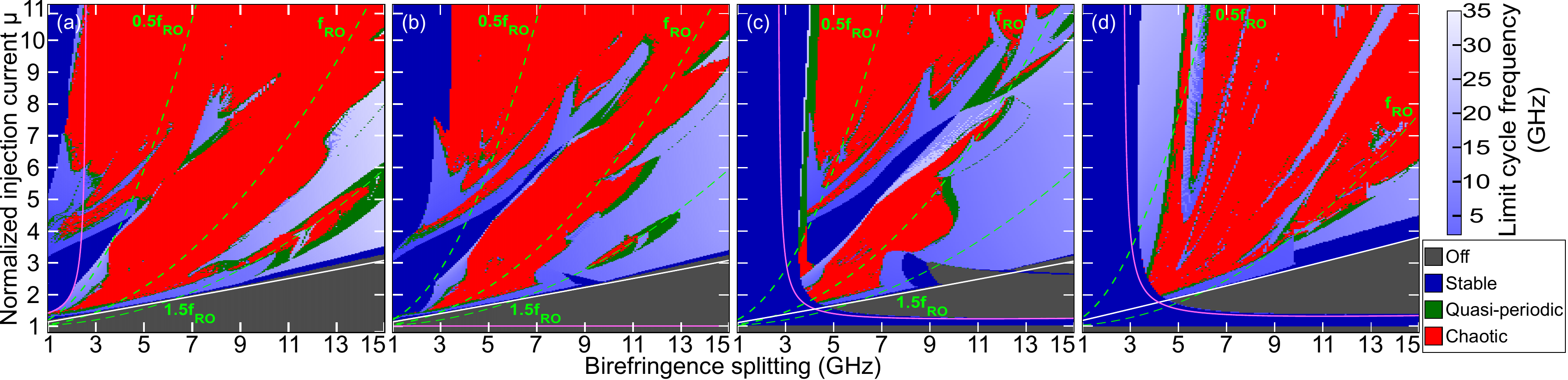}
    \caption{Dynamics maps for different values of the injection current $\mu$ and the birefringence splitting $f_{p,0}$. The dynamics of the multimode model are displayed for (a) $\gamma_{a}$ = -1 ns$^{-1}$, (b) $\gamma_{a}$ = 0 ns$^{-1}$ and (c) $\gamma_{a}$ = 1 ns$^{-1}$. (d) The dynamics of the single-mode model is displayed for $\gamma_{a}$ = 1 ns$^{-1}$. The same color code as in Fig.~\ref{fig:diag_bifur_spectre_RF} is used. The periodic states are indicated in light blue, where the gradient of blue quantifies the value of the dominant frequency. The green dashed lines indicate $f_{RO}/2$, $f_{RO}$ and $3 f_{RO}/2$. The white and magenta curves respectively represent the stability boundaries of the LP$_{01,x}$ and the LP$_{01,y}$ modes as computed from the Hopf bifurcation expressions $\mu_{x,H}$ and $\mu_{y,H}$ detailed in the text.}
    \label{fig:map_dyna}
\end{figure*}

To analyze the temporal dynamics, the bifurcation diagram of the LP$_{01,y}$ mode is plotted in Fig.~\ref{fig:diag_bifur_spectre_RF}(b), and the corresponding RF-spectra in Fig.~\ref{fig:diag_bifur_spectre_RF}(c). Firstly, as the current increases, the LP$_{01,y}$ mode bifurcates from a static state to different dynamics such as periodic, quasi-periodic and even chaotic dynamics. An example of a quasi-periodic dynamics is shown in Fig.~\ref{fig:trace_spectre_quasi_per_chaos}(A) for the two LP$_{01}$ modes. While a bifurcation to chaos from periodic or quasi-periodic dynamics is not surprising \cite{virte2013bifurcation}, it is unexpected for the dynamics to return to a stable regime at $\mu \approx 7.4$ after a chaotic regime. In particular, we have checked that the simulation for a SM-VCSEL does not feature such a restabilization [see Fig.~\ref{fig:map_dyna}(d)]. Moreover, the chaotic regimes in different current ranges show different levels of instability, quantified by the Lyapunov exponent \cite{wolf1985determining}. For the three current intervals featuring chaos, we find maximum Lyapunov exponents (averaged over each interval) of approximately 0.04 ns$^{-1}$, 0.05 ns$^{-1}$ and 0.08 ns$^{-1}$, respectively. This highlights that the third chaotic region exhibits an increased instability compared to the other two. % paragraph

% figure 2

Figure~\ref{fig:trace_spectre_quasi_per_chaos}(B) shows a chaotic time trace and a RF-spectrum for the LP$_{01}$ mode at $\mu = 6.20$. As expected, the time trace is complex and unpredictable. However, while the RF-spectrum is fairly broad, the presence of dominant frequency components around the birefringence splitting $f_{p,0}$ and its harmonics indicates that the self-pulsation originates from the polarization dynamics. While a few prominent peaks remain in the RF-spectrum, the difference between the chaotic and the quasi-periodic spectrum is clear. Furthermore, Fig.~\ref{fig:diag_bifur_spectre_RF}(c) reveals that the chaotic dynamics is caused by the interplay between the relaxation oscillation frequency (ROF) $f_{RO} = \sqrt{2\kappa\gamma_{e}(\mu-1)}/(2\pi)$ and the birefringence splitting $f_{p,0}$ which reach similar values in this current range \cite{virte2013bifurcation}. Indeed, by comparing Fig.~\ref{fig:trace_spectre_quasi_per_chaos}(A.2) and Fig.~\ref{fig:trace_spectre_quasi_per_chaos}(B.2), we notice that for $\mu$ = 6.20 the main frequency shifted a bit from $f_{p,0}$ towards $f_{R0} \approx 10.4$~GHz. % paragraph

In the RF-spectrum in Fig.~\ref{fig:diag_bifur_spectre_RF}(c), $f_{p,0}$ and its harmonics are clearly observed. Furthermore, non-commensurable frequencies appear in quasi-periodic regimes, and sub-harmonics in periodic regimes. Interestingly, both the bifurcation scenario leading to PS, subsequent quasi-periodic and then chaotic dynamics, and the dominant frequencies characterizing the dynamics, are qualitatively similar to recent experimental studies of MM-VCSELs \cite{Bittner2022, mercadier2025chaos}. Hence, while only two TMs are simulated, it appears that the model qualitatively reproduces dynamics observed experimentally for MM-VCSELs with up to 12 TMs \cite{Bittner2022, mercadier2025chaos}. % paragraph

\begin{figure}[t] 
    \centering
    \includegraphics[width=\linewidth]{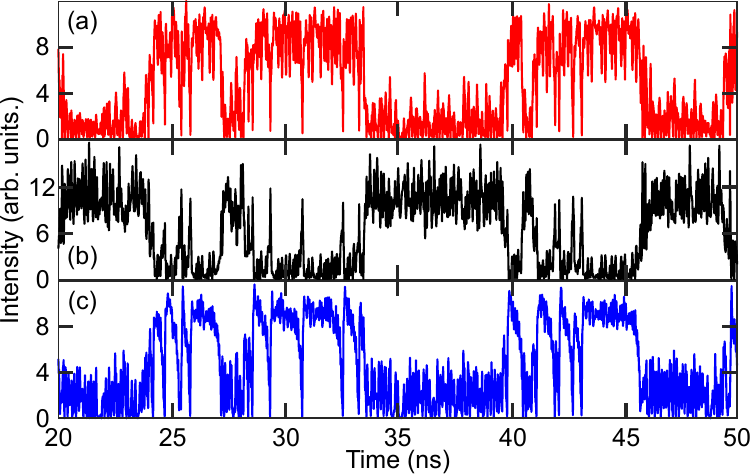}
    \caption{(a) Temporal traces of the LP$_{01,x}$, (b) the LP$_{01,y}$ and (c)~the LP$_{11,y}$ modes for $\mu = 11$, filtered by a Butterworth low-pass filter of order 4 with cutoff frequency $f_{p,0} = 9.7$~GHz. Other parameters as in Fig.~\ref{fig:diag_bifur_spectre_RF}.}
    \label{fig:polarization_transverse_modes_switching}
\end{figure}

Lastly, it is interesting to explore the dynamics in the third chaotic zone. When analyzing the bifurcation diagram in figure \ref{fig:diag_bifur_spectre_RF}(b), two zones can be distinguished in this chaotic region with a separation at $\mu \approx 10.2$. This separation can be clearly observed in the RF-spectrum, which exhibits high amplitudes for low frequencies for $\mu \gtrsim 10.2$, indicating a slow dynamical process. In order to understand the dynamics for this current range, the temporal traces of the LP$_{01,x}$ and LP$_{01,y}$ modes are plotted in Figs.~\ref{fig:polarization_transverse_modes_switching}(a) and (b). A Butterworth low-pass filter of order 4 with a cutoff frequency of $f_{p,0}$ was applied to better reveal the slow dynamics. The strong anticorrelation of the time traces indicates polarization-mode hopping (PMH), a phenomenon known for SM-VCSELs \cite{virte2013deterministic}. The temporal trace of the LP$_{11,y}$ mode in Fig.~\ref{fig:polarization_transverse_modes_switching}(c) is also highly anticorrelated with LP$_{01,y}$, an effect we call transverse-mode hopping (TMH). TMH can evidently not exist in a SM-VCSEL and has not yet been observed owing to the difficulty to spectrally isolate the time traces of different modes experimentally. The time trace of LP$_{11,x}$ (not shown) is also anticorrelated to LP$_{11,y}$ (PMH) and to LP$_{01,x}$ (THM), so polarization- and transverse-mode hopping are linked. Furthermore, to determine whether this dynamics is driven by noise or chaos, the evolution of the dwell time as a function of the injection current was analyzed. The dwell time decreases on average with the current, which cannot be explained by a noise-induced Kramers hopping problem \cite{willemsen1999polarization}, but is the consequence of deterministic chaotic dynamics \cite{olejniczak2011polarization}. % paragraph

% figure 3

Next, we modify the amplitude anisotropy $\gamma_{a}$ and map the dynamics as function of current and birefringence splitting $f_{p,0}$. The results in Figs.~\ref{fig:map_dyna}(a)--(c) are for the MM-VCSEL with $\gamma_{a}$ = -1, 0, and 1~ns$^{-1}$, and for the SM-VCSEL model with $\gamma_{a}$ = 1 ns$^{-1}$ in Fig.~\ref{fig:map_dyna}(d). Comparing Figs.~\ref{fig:map_dyna}(a)--(c) shows the impact of the amplitude anisotropy. When decreasing $\gamma_{a}$, a chaotic area for low currents and high birefringence splittings appears. In contrast, when increasing $\gamma_{a}$, an area of stable dynamics between the chaotic zones appears and grows. In general, decreasing $\gamma_{a}$ seems to facilitate chaotic dynamics. Moreover, comparing Figs.~\ref{fig:map_dyna}(c) and (d) shows the differences between SM- and MM-VCSELs. As mentioned above, the two cases produce similar results for low currents. We plot in Fig.~\ref{fig:map_dyna} the stability boundaries for the LP$_{01,x}$ and LP$_{01,y}$ modes of a SM-VCSEL, as approximated by the Hopf bifurcation conditions $\mu_{x,H} = 1 + \frac{\gamma_{s}}{\kappa\alpha-\gamma_{p}}\frac{\gamma_{p}}{\gamma_{e}}$ and $\mu_{y,H} = 1 + \frac{2(\gamma_{s}^2+4\gamma_{p}^2)}{\kappa(2\alpha\gamma_{p}-\gamma_{s})}\frac{\gamma_{a}}{\gamma_{e}}$ \cite{Panajotov2013a}. Interestingly, these stability boundaries also predict the dynamics of a MM-VCSEL at low currents. However, once the LP$_{11}$ modes are lasing, the dynamics of SM- and MM-VCSELs differ. In particular, even if adding another TM should make the MM-VCSEL more complex than the SM-VCSEL, the dynamics may instead be more stable with the onset of periodic or even static dynamics after going through chaotic regimes, which is found for all three values of $\gamma_a$ in Fig.~\ref{fig:map_dyna}. Indeed, the large area of stable dynamics for the MM-VCSEL cannot be observed in the SFM for SM-VCSELs. It is also worth noting that, while the dynamics for $f_{p,0} \lesssim 3.5$~GHz is stable for all pump currents in Fig.~\ref{fig:map_dyna}(c), it is possible to obtain a more complex dynamics by decreasing the value of the spin relaxation rate $\gamma_{s}$. In contrast, when the spin relaxation rate $\gamma_{s}$ is above $\gamma_{s,th} \approx$ 125 ns$^{-1}$, the system no longer becomes chaotic for this current range. When $\gamma_{s}$ increases, the interactions between the two carrier reservoirs are so fast that they become basically equal, removing the nonlinear spin-induced carrier coupling and preventing the onset of chaos in the framework of the SFM. % paragraph

Furthermore, the maps indicate the main frequency of the limit cycles in periodic regimes. As anticipated, chaos seems to appear when the limit cycle frequency ($\approx f_{p,0}$) is close to $m \, f_{RO}$ with $m \in\{0.5,1,1.5\}$. Figure~\ref{fig:map_dyna}(c) shows another surprising effect for $\mu \approx 5.5$ and $f_{p,0} \approx 8.5$~GHz: the limit cycle frequency is very large around 2$f_{p,0}$ after leaving the chaotic regime, and suddenly drops to about $f_{p,0}$ when increasing the current further. % paragraph

In summary, our analysis of the dynamics of a MM-VCSEL shows that increasing the current leads to successive switchings between modes with different polarization and transverse order. This mode competition results in various dynamical regimes including chaos. Our simulations also highlight the role of the birefringence splitting and the ROF in the dynamics. This is in qualitative agreement with MM-VCSEL experiments showing a complex evolution of the polarization state and TM configuration with the pump current and multiple regimes of chaotic dynamics \cite{Bittner2022, mercadier2025chaos}. So the SFM is a good model for the dynamics of free-running MM-VCSELs, and key features of their dynamics are found even for a model with only two TMs. The Lyapunov exponents in the chaotic regimes are 0.04--0.08 ns$^{-1}$, so the level of instability can be adjusted via the current. Finally, our comparison of MM- and SM-VCSELs highlights two novel phenomena: first, the presence of the LP$_{11}$ mode favors a return to static dynamics after a chaotic one, which is not found in our simulations for SM-VCSELs. Second, a new chaotic dynamics involving both polarization and transverse mode hopping is found.

\acknowledgments{The Chair in Photonics is supported by Region Grand Est, GDI Simulation, Departement de la Moselle, European Regional Development Fund, CentraleSupelec, Fondation CentraleSupelec, and Eurometropole de Metz. A.V.\, acknowledges funding from Ministerio de Ciencia e Innovación (Spain), (PID2021-123459OB-C22 MCIN/AEI/FEDER,UE).}

\newpage
% Bibliography
%\bibliography{Refs}


\end{document}